\documentstyle[twocolumn,prb,aps]{revtex}
\begin{document}
\twocolumn[\hsize\textwidth\columnwidth\hsize\csname 
@twocolumnfalse\endcsname
\date{\today}
\draft
\title{\bf
Topological Phase Transition in the
$\nu = 2/3$ Quantum Hall Effect}
\author{I. A. McDonald~\cite{byline}$^{(1),(2)}$
and F. D. M. Haldane$^{(1)}$}
\address{
$^{(1)}$ Department of Physics, Princeton University\\
Princeton, New Jersey 08540 \\
$^{(2)}$ Department of Physics, Penn. State University\\
University Park, PA 16802}
\maketitle
\begin{abstract}
The double layer $\nu=2/3$
fractional quantum Hall system is studied using
the edge state formalism 
and finite-size diagonalization subject to periodic
boundary conditions.
Transitions between three different ground states
are observed as the separation 
as well as the tunneling between
the two layers is varied.  
Experimental consequences are discussed.
\end{abstract}
\pacs{PACS numbers: 2.60.Cb 73.20.Dx 73.40.Hm}
]
\narrowtext
\section{Introduction}
Advances in semiconductor fabrication have made it
possible to produce multilayer two dimensional
electron systems that allow exploration of the
effects of interlayer as well as intralayer 
correlations.  The possibilities raised by
the introduction of
an extra degree of freedom into the
standard picture of the fractional
quantum Hall effect were first examined
by Halperin~\cite{Halp1} in the context
of spin-unpolarized ground states,
and later by Haldane and Rezayi~\cite{Hald2}
who proposed applying Halperin's wavefunctions
to the case where the electrons
possess a
double valued index indicating the quantum state of the
electron in the third direction parallel
to the magnetic field.  These states 
have been analyzed in their spin as well as
layer index form~\cite{Yosh}.
Recent experiments with
multilayer electron systems~\cite{Eisen1,Suen} seem 
to suggest the existence of incompressible
states which belong to the universality class
described by Halperin's wavefunctions.
These systems allow the manipulation of two
parameters which greatly influence the 
character of the fluids, 
the distance between the
effective layers of electrons, given by $d$ and
the tunneling between the two layers, denoted
by the energy difference between the lowest two
subbands as $\Delta_{sas}.$

Motivated by the experimental observation of a transition
between two distinct $\nu = 2/3$ states in wide 
quantum wells by Suen, et.al.~\cite{Suen} and also
the transition between spin-polarized and
spin-singlet $\nu =2/3$ states in tilted field experiments
by Eisenstein, et.al.~\cite{Eisen2},
we have analyzed the double-layer model 
with interlayer tunneling at total filling factor
$\nu =2/3$ using the edge state formalism and
a finite-size numerical 
diagonalization study. 
In the double layer model electrons are localized 
on a pair of parallel planes, between which they can tunnel.  
Tunneling lowers the energy of the electrons in 
a symmetric combination of states in each plane
relative to the anti-symmetric state.
If symmetric states are
identified as pseudospin ``up'' states,
and anti-symmetric states as pseudospin
``down'', tunneling acts as a Zeeman coupling in
pseudospin space.  Therefore our investigation of the 
double layer model will allow a very general picture to emerge
of two component systems where the pseudospin can refer
either to real spin or to subband index.

We have found a rich phase diagram with three distinct
phases, separated by what we identify as first-order
transitions:
(1) The particle-hole conjugate of the
$\nu =1/3$ Laughlin state, where all the 
electrons are in the symmetric state of the double layer,
(2) The pseudo-spin singlet analog
of the $\nu =2/3$ spin-singlet state,
where the electrons are divided equally between 
symmetric and antisymmetric states, and
(3) a state with independent $\nu =1/3$ Laughlin 
states in each layer.
An interesting feature is that states
(1) and (2), while distinct, have been 
identified as having the same topological
order~\cite{Wen1}, different from that
of (3).  We carried out our investigation in the
periodic (or toroidal) geometry, which is well
adapted to exhibit these differences.
For example, the long distance
effective Chern-Simons (CS) theory
of the Hall effect predicts a three-fold
ground state degeneracy for states (1) and (2)
in the periodic geometry, but a nine-fold
degeneracy for state (3),
consistent with our observation.
An interesting consequence of the
fact that state (3) has different topological order
from states (1) and (2) is that the boundary between
state (3) and either state (1) or (2) necessarily
supports a residual neutral gapless Luttinger 
liquid~\cite{Hald6}.
The coexistence of two, and possibly three phases of
quantum Hall states at $\nu =2/3$ raises many 
interesting questions, both experimental and
theoretical.
 
\section{Effective Theories and Edge States}
Rather than work with the wavefunctions
themselves, we can alternatively discuss
the fractional quantum Hall states in
terms of the long-wavelength effective field theories
describing the incompressible fluids.
These low-energy effective theories capture
the long-distance correlations between
the particles, determining such universal
quantities as the conductance and the
charge and statistics of the quasiparticles.
The Lagrangian density for the effective theory
can be written in the form
\begin{equation}
2\pi{\cal L} = \epsilon^{\lambda\mu\nu} 
\left (\mbox{$\case{1}{2}$}
\hbar ({\bf a}_{\lambda},{\bf K}\partial_{\mu}
{\bf a}_{\nu}) + eA_{\lambda} 
({\bf q},\partial_{\mu} {\bf a}_{\nu}) \right ),
\label{cslagrangian}
\end{equation}
where ${\bf a}_{\mu}$ is an $n$-component vector 
of Abelian CS gauge fields,
$A_{\mu}$ is the electromagnetic gauge field, 
${\bf K}$ is a nonsingular
integer coupling matrix,
${\bf q}$ is an integer vector; 
$({\bf a},{\bf b})$ is the inner product.
We note here that we
use the notation ${\bf q}$
for the charge vector rather than
${\bf t}$ as used by Wen and Zee~\cite{Wen1}.
By integrating out the Chern-Simons fields
${\bf a}_{\nu}$ we can determine the Hall 
conductance of the above theory to be
\begin{equation}
{\sigma}^H = {e^2 \over h} ({\bf q},{\bf K}^{-1} {\bf q}).
\end{equation}
The effective theory allows vortex defects
of the Chern-Simons fields, with
core energies determined by short-distance
terms in the Hamiltonian which haven't been
included.
One can determine the charge and
statistics of a particular composite
of vortices by specifying 
an integer valued vector ${\bf l}$
such that the composite in question
is made up 
of $l_i$ vortices of type $i$.
The charge of this composite is
then given by
\begin{equation}
Q = e ({\bf q},{\bf K}^{-1} {\bf l})
\end{equation}
and the statistical phase acquired when two
such composite  
vortices are interchanged is given by
\begin{equation}
{{\theta} \over \pi} = ({\bf l},{\bf K}^{-1} {\bf l}).
\end{equation}
Furthermore, it has been shown~\cite{Wen1}
that the degeneracy of a state described by
the matrix $K$ when it is 
defined on a two-dimensional closed
surface of genus $g$ is given by
\begin{equation}
{\bf D} = {\vert {\rm Det} {\bf K} \vert}^{g}.
\end{equation}
We may therefore classify
an Abelian quantum Hall state
by specifying an integer
valued pair 
$\{ {\bf K},{\bf q} \}$~\cite{Wen1},
thereby determining the long-distance properties
of the fluid.
It is important to 
note that distinct quantum Hall states
are represented by equivalence classes
of $\{ {\bf K},{\bf q} \}$ pairs as the
above properties are invariant under 
$SL({\kappa},Z)$ basis changes
\begin{eqnarray}
{\bf K} & \rightarrow {\bf W} {\bf K} {\bf W}^{T} \nonumber\\
{\bf q} & \rightarrow {\bf W} {\bf q}
\end{eqnarray}
where 
${\bf W}$ is an integer matrix with
$ \vert {\rm Det} {\bf W} \vert = 1.$
Two fractional quantum Hall states described
by $\{ {\bf K},{\bf q} \}$ pairs which
are related by an 
$SL({\kappa},Z)$ transform belong to the
same universality class and are considered
topologically equivalent.

The above discussion has been limited to
Abelian fractional
quantum Hall fluids on the plane.  If we instead define our
effective theory on a sphere, there is an extra
term in the effective theory which describes 
the coupling of the liquid to the curvature
of the sphere.  This manifests itself in
a new topological quantum number, the
flux shift,
which is not determined by the long distance
effective theory described above.
The shift ${\cal S}$,
defined by
the relation
$N_{\phi} = {\nu}^{-1} N_e - {\cal S}$,
is a manifestation of the coupling of the
orbital angular momentum properties of the
state to the curvature of the space. 
The effective theories
employed in the
composite fermion approach on the
plane do not distinguish between 
Landau levels and spin states.
As the orbital angular momentum carried by
by the cyclotron motion of electrons
in the second Landau level is different
from electrons in the lowest level,
the shift ${\cal S}$ provides a way
to distinguish between states that
possess the same long distance properties
but which have different spin symmetries.
By utilizing the topological character of the
coupling of orbital 
angular momentum to curvature
we can classify the spin structure of a
state whose analysis
would otherwise lie outside
the effective theory approach. 

We can write the 
Halperin-Laughlin wavefunction
appropriate for multi-component 
systems in the planar geometry in the form
\begin{equation}
\Psi_{\bf K}(\{ z_i \}) 
= \prod_{i<j} (z_i - z_j)^{K^{el}({\sigma}_i,{\sigma}_j)}
\prod_i {\rm exp} \biggl(-{1 \over 4 l^2} 
{\vert z_i \vert}^2 \biggr)
\end{equation}
where $\sigma_i$ is the pseudospin
variable and
$K^{el}_{ij} = K^{el}({\sigma_i},{\sigma_j})$
is the symmetric matrix encoding the electron
correlations.
If we specialize to the double layer
system, we can identify the $(m_1,m_2,n)$
state with the effective theory
\begin{equation}
{\bf K} = {\bf K}^{el} = \left (
\begin{array}{cc}
m_1 & n \\
n & m_2
\end{array}
\right )
\ \
{\bf q} = \left (
\begin{array} {cc}
1 \\
1
\end{array}
\right ).
\end{equation}

We shall be considering two effective theories
for the system at $\nu=2/3$.
The first has been identified as representing
both the pseudospin singlet state and the pseudospin
polarized, particle-hole conjugate of
a Laughlin $\nu=1/3$ state.
The effective theory is given by
\begin{equation}
{\bf K} = \left (
\begin{array}{cc}
1 & 2 \\
2 & 1
\end{array}
\right )
\ \
{\bf q} = \left (
\begin{array} {cc}
1 \\
1
\end{array}
\right ).
\end{equation}
The second effective theory that we will consider
represents two independent $\nu =1/3$ Laughlin
states
\begin{equation}
{\bf K} = \left (
\begin{array}{cc}
3 & 0 \\
0 & 3
\end{array}
\right )
\ \
{\bf q} = \left (
\begin{array} {cc}
1 \\
1
\end{array}
\right ).
\end{equation}
Both effective theories potentially represent
states at $\nu=2/3$.  There is however a crucial
difference between the two theories:
they possess different ground state degeneracies
on a non-trivial closed space,
and therefore possess different topological order.
In the remainder of this paper we will
investigate the consequences of the
fact that the these two theories have different
topological order.
 
Let us consider the edge state theory 
of the Abelian quantum Hall states.
The edge between Abelian Hall fluids
is a one-dimensional ``Luttinger liquid''
which can also be characterized by
a $\{ {\bf K},{\bf q} \}$ pair, which is
the same pair of the bulk theory if the
edge is between a Hall state and a 
non-conducting state.
Generally, we may 
define a set of fields $\phi_i (x)$
living on the one-dimensional
compact edge of an
incompressible Hall sample such
that they obey the equal time
commutation relations
\begin{equation}
[{\phi}_i (x), {\phi}_j (x^{\prime})] = i \pi
(K_{ij} {\rm sgn}(x-x^{\prime}) + L_{ij})
\end{equation}
where $L_{ij} = {\rm sgn} (i-j) (K_{ij} + q_i q_j)$
is a Klein factor.
The action for the 
{\it translationally invariant}
edge fields is given by
\begin{equation}
S = \int dt \  (S_0 - H_0)
\end{equation}
where
\begin{equation}
S_0 = - {\hbar \over 4 \pi} \sum_{ij} K_{ij}^{-1}
\oint dx\  \partial_x \phi_i \partial_t \phi_j
\end{equation}
and
\begin{equation}
H_0 = {\hbar \over 4 \pi} \sum_{ij} V_{ij} 
\oint dx\ \partial_x \phi_i \partial_x \phi_j
\end{equation}
where $V_{ij}$ encodes the non-universal
interactions which determine,
among other things, the velocities
of the various edge modes.
For stability, we require the
matrix ${\bf V}$ to be positive definite.
Let us define 
\begin{equation}
{\phi}_{\bf m} = \sum_i m_i {\phi}_i
\end{equation}
and
\begin{equation}
q({\bf m}) = ({\bf m},{\bf q}) 
\end{equation}
where
${\bf m}$ is an integer valued vector.
We can define a set of composite 
local fields
\begin{equation}
\Psi_{\bf m} = e^{-i \phi_{\bf m}}
\end{equation}
which obey, for $x \not= x^{\prime}$
\begin{equation}
\Psi_{\bf m} (x) \Psi_{{\bf m}^{\prime}} (x^{\prime})
= \eta_{{\bf m},{\bf m}^{\prime}} 
\Psi_{{\bf m}^{\prime}} (x^{\prime}) \Psi_{\bf m} (x)
\end{equation}
where
\begin{equation}
\eta_{{\bf m},{\bf m}^{\prime}} 
= (-1)^{q({\bf m}) q({\bf m}^{\prime})}.
\end{equation}
The winding number operator, 
defined as
\begin{equation}
N_i = {1 \over 2 \pi} \oint dx \ \partial_x \phi_i (x)
\end{equation}
is constrained to be integral by
imposing 
periodic boundary conditions on the fields
\begin{equation}
\Psi_{\bf m} (x+L) = \Psi_{\bf m} (x).
\end{equation}
The fields are characterized by the integer
quadratic form
\begin{equation}
K({\bf m}) = ({\bf m},{\bf K} {\bf m}) = \sum_{ij} m_i K_{ij} m_j
\end{equation}
which has the constraint
\begin{equation}
(-1)^{K({\bf m})} = (-1)^{q({\bf m})}.
\end{equation}
If $K({\bf m})$ is odd, the field
corresponding to ${\bf m}$ is
is fermionic, and if $K({\bf m})$
is even, it is bosonic.
We can define the charge density of the
edge as
\begin{equation}
\rho(x) = {e \over 2 \pi} \sum_{ij}
q_{i} {K_{ij}^{-1}} \partial_x
\phi_{j} (x)
\end{equation}
which has the
commutation relation
\begin{equation}
[\rho(x),\rho(x^{\prime})] = i \hbar \sigma^{H}
\delta^{\prime} (x-x^{\prime})
\end{equation}
where
\begin{equation}
\sigma^{H} = {e^2 \over h} \sum_{ij}
q_{i} K_{ij}^{-1} q_{j}.
\end{equation}
The physical interpretation of this
quantity is that it
represents 
the change in the
quantized Hall conductivity in going
from one side of the edge to the other.
We can also define the total charge
operator 
\begin{equation}
Q = e \sum_{ij} K_{ij}^{-1} q_i N_{j} 
\end{equation}
which obeys the commutation relation
\begin{equation}
[Q,\Psi_{\bf m}(x)] = e q({\bf m}) \Psi_{\bf m} (x).
\end{equation}

The edge state Hamiltonian $H_0$
describes $n$ linearly independent
oscillator 
modes propagating with velocities $v_{\lambda}$
which can be determined from
the generalized real symmetric eigenvalue
equation
\begin{equation}
{\bf V} {\bf u}_{\lambda} = v_{\lambda} {\bf K}^{-1}
{\bf u}_{\lambda}
\end{equation}
where the velocities are real and
the eigenmodes independent as ${\bf V}$
is positive definite.
We can now state the condition on the matrix
${\bf K}$ which ensures a set of $SU(2)$
generating operators within the theory:
If we can identify an integral
vector ${\bf m}$ such that
\begin{equation}
K({\bf m}) = 2 s_{\bf m}
\ \ \ q({\bf m}) = 0
\label{big}
\end{equation}
where $s_{\bf m} = \pm 1$
then we may 
identify an $SU(2)$ algebra
associated with the
edge field $\Psi_{\bf m}.$
We can
define a triplet of non-Abelian densities
\begin{eqnarray}
{\sigma^x} (x) &= {1 \over 2} \bigl(\Psi_{\bf m}(x) 
+ \Psi_{-{\bf m}} (x)\bigr)
\nonumber\\
{\sigma^y} (x) &= {s_{\bf m} \over 2 i} 
\bigl( \Psi_{\bf m} (x)- \Psi_{-{\bf m}}(x) \bigr)
\nonumber\\
{\sigma^z} (x) &= {1 \over 2} \rho_{\bf m} (x) =
{1 \over 4 \pi} \partial_x {\phi}_{\bf m} (x)
\end{eqnarray}
which obey a level-1 $SU(2)$ Kac-Moody
algebra
\begin{eqnarray}
[\sigma^{a}(x),\sigma^{b} (x^{\prime})]
&=& {i s_{\bf m} \over 4 \pi} \delta^{\prime} 
(x-x^{\prime}) 
\nonumber\\
&+& i \epsilon^{a b c} \sigma^{c} (x)
\delta (x-x^{\prime}).
\end{eqnarray}
We may identify the $SU(2)$ algebra generators
\begin{equation}
S^{a} = \oint dx\ \sigma_{\bf m}^{a} (x)
\label{alg}
\end{equation}
which obey the commutation relations
\begin{equation}
[S^{a},S^{b}] = i \epsilon^{a b c} S^{c}.
\end{equation}
The auxiliary constraint $q({\bf m})=0$
is to insure charge neutrality. 
This hidden $SU(2)$ symmetry has been
noted previously~\cite{Frohlich,Kane,Wil}.
We note that larger symmetries (specifically $SU(N)$
for an N-component system)
may be identified using a similar analysis.

Let us focus specifically on the 
double component system,
where the vector
components denote different pseudospin
components.  We shall work in
the symmetric basis where
$q_i =1$ for $i=1,2$.
Further, the diagonal elements 
$K_{ii}$ must be odd for fermi statistics.
We therefore find that the only 
effective theory which
may generate an
$SU(2)$ edge symmetry is
\begin{equation}
{\bf K} = \left(
\begin{array} {cc}
m\pm 1 & m \\
m & m \pm 1 \\
\end{array}
\right)
\label{sps}
\end{equation}
with $m$ even.
Diagonalizing the above edge state system~(\ref{sps})
we obtain the Hamiltonian
\begin{eqnarray}
H = {\hbar \over 4 \pi} \oint dx \ (v_n (\partial_x \phi_{n} (x))^2
+ v_c (\partial_x \phi_{c}(x))^2  \nonumber\\
+ 2 v_{int} \partial_x  \phi_c (x)  
\partial_x \phi_n (x))
\end{eqnarray}
where
$\phi_{n} = \phi_1 - \phi_2$
is a neutral mode and
and $\phi_c = \phi_1 + \phi_2$
is a charged mode.
The two modes move 
with velocities
that depend on the external potential,
with intermode interactions $v_{int}$ 
mixing these two modes.
It is the neutral
edge mode that generates the $SU(2)$ algebra
where the field
\begin{equation}
\Psi_{\bf m} = e^{-i(\phi_1 - \phi_2)}
\end{equation}
corresponds to the neutral physical operation
of tunneling one electron from
one pseudospin to the other.
The
effective theory defined by~(\ref{sps})
possesses a hidden $SU(2)$ 
symmetry of the bulk quantum state,
in this case invariance under pseudospin
rotations.
We may therefore state that a bulk
quantum Hall state may possess an
$SU(2)$ symmetry if and only if the
effective theory corresponding to this quantum
Hall state has a non-trivial
solution ${\bf m}$ to the equation
$K({\bf m}) = \pm 2.$ 
As the bulk ground state 
is non-degenerate, apart from topological
degeneracies which are not associated
with the $SU(2)$ symmetry, our states
constructed in this way are $SU(2)$ singlets.
Whether or not a particular
state is realized as the
ground state depends, of course,
on the underlying bulk Hamiltonian.
It is important to realize that this
symmetry is a consequence of the electron
correlations rather than the underlying 
Hamiltonian.  This algebra is realized in
the bulk state independently of whether or not
the Hamiltonian is strictly
invariant under the symmetry.

We must note that in our construction
we have assumed that the matrix ${\bf K}$
was non-singular.  The singular case
of $K_{ij} = m$ with $m$ odd corresponds 
to a $\nu = 1/m$ state which is also
invariant under pseudospin rotations, but
which is fully polarized with $S = N_e /2.$
This system has been studied previously
and is found 
to possess many interesting features~\cite{Macd}.

The edge state formalism is also capable of
addressing the question of what happens 
at the edge between two Abelian Hall states,
which is physically realized when there is
phase coexistence in a first-order transition
between the two states.
We can form the edge state
theory corresponding to 
the $\{ {\bf K},{\bf q} \}$ effective
by forming the direct sum
${\bf K} = {\bf K}_1 \oplus -{\bf K}_2$
and ${\bf q} = {\bf q}_1 \oplus {\bf q}_2$.
We will restrict our discussion to a {\it clean}
edge between
two 
double component 
Hall states at the same
filling fraction, 
at least one of which possesses an
$SU(2)$ symmetry of the type 
discussed above. 
To address the
stability of the edge of this system,
we must consider the general 
non-winding number conserving tunneling
perturbation
\begin{equation}
T = \int dx \bigl[ t(x) \Psi_{{\bf m}} (x) + c.c. \bigl].
\end{equation}
The operator must be bosonic
and charge conserving with
$q({\bf m}) = 0$ and
$K({\bf m})$ even.
The scaling dimension of this
operator can be determined from
the two-point function to be
$2 - \Delta({\bf m})$,
where
$\Delta({\bf m})$ satisfies the
inequality
\begin{equation}
\Delta({\bf m}) \ge {1 \over 2} \vert K({\bf m}) \vert.
\end{equation}
We can construct a representation of
the fields such that
both $K_{ij}$ and the non-universal
$V_{ij}$ are diagonal
\begin{equation}
(U K U^{T})_{ij} = \sigma_i \delta_{ij}
\end{equation}
\begin{equation}
(U V U^{T})_{ij} = {\hat v}_i \delta_{ij}
\end{equation}
where $\sigma_i = \pm 1.$
Since stability requires ${\hat v}_i > 0$,
the direction of propagation of each mode is
determined by $\sigma_i$.
Using the transformation matrix $U$ we
can determine $\Delta({\bf m})$
\begin{equation}
\sum_i \eta_i({\bf m}) = 2 \Delta({\bf m})
\end{equation}
where 
\begin{equation}
\eta_i({\bf m}) = \sum_{jk} U^{-1}_{ji} U^{-1}_{ki} m_j m_k.
\end{equation}
The factors 
$\eta({\bf m})$ obey the
sum rule
\begin{equation}
\sum_i \eta_i({\bf m}) \sigma_i = K({\bf m}).
\end{equation}
One can see that if the system is maximally chiral, then
the sum rule explicitly gives us the scaling dimension
of the operator.  If it is not maximally chiral, it
only gives us a lower bound.
Therefore, from a scaling
perspective a tunneling operator with
$K({\bf m}) = 0$ is potentially relevant 
if the scaling dimension 
$\Delta({\bf m}) < 2$.
In general, the tunneling perturbation will
be prevented from being relevant by the 
complex tunneling parameter $t(x)$ where
\begin{equation}
t(x) = \vert t(x) \vert {\rm exp} [i \alpha (x)]
\end{equation}
in the clean case.
In order for the perturbation to be relevant, the
phase factor must satisfy
\begin{equation}
\partial_x \alpha (x)
= \langle 0 \vert \partial_x \phi_{\bf m} \vert 0 \rangle.
\end{equation}
We will assume that such a ``phase locking''
is generically possible for the edges in
question.
If the $K({\bf m})=0$ perturbation is
relevant, the modes involved in $\Psi_{\bf m}$
become massive and are removed from the low-energy 
theory.  
The only other perturbations that can be potentially
relevant have $\vert K({\bf m}) \vert = 2$ but are not
mass generating (they do form 'hidden' $SU(2)$
symmetries as discussed previously).  Therefore, the condition for
a potentially mass generating perturbation is
that we must 
identify a non-trivial integer valued vector
${\bf m}$ such that $K({\bf m}) = q({\bf m}) = 0.$ 
We shall take the point
of view in this paper that if a mass-generating 
perturbation is potentially relevant, the system will
relax so that the instability generally occurs.
This is a conjecture based upon the experimental
observation that the only stable Abelian Hall states 
are those that do not permit such 
mass-generating instabilities~\cite{Hald6}. 
If an mass-generating instability can occur, 
it appears to do so.

We can apply this analysis to the edge theory at hand.
First, let us consider the edge between two states with
the same topological structure.
The edge state theory will be based
on the $\{ {\bf K},{\bf q} \}$ pair
\begin{equation}
{\bf K} = \left(
\begin{array} {cccc}
m\pm 1 & m & 0 & 0 \\
m & m \pm 1 & 0 & 0\\
0 & 0 & -(m \pm 1) & -m \\
0 & 0 & -m & -(m \pm 1)\\
\end{array}
\right)
\end{equation}
with
\begin{equation}
{\bf q} = \left(
\begin{array} {c}
1 \\
1 \\
1 \\
1 \\
\end{array}
\right).
\label{edge1}
\end{equation}
We note that the above edge
has $\sigma_H = 0$ as is appropriate for
an edge between two Hall states with the
same filling fraction. 
In this case, we can identify the
two operators of interest: the operator
$\Psi_{\bf m}$ 
with ${\bf m} = (1,0,-1,0)$
and the operator with
${\bf m} = (0,1,0,-1)$.
Both these operators are $K({\bf m})=0$ operators which are potentially
relevant and mass-generating.  Interactions will tend to
reduce the scaling dimension of the two operators,
but if they remain relevant we expect that
they will cause a mass gap to form (we note
that generally diagonal
elements of the interaction matrix 
reduce the scaling dimension of
the operators, while off-diagonal elements 
tend to increase it, upto a maximum
of $2-{1 \over 2} \vert K({\bf m}) \vert$).  It is straightforward
to see that any edge between states with the same
$K$ matrix will have two such tunneling operators
with $K({\bf m}) = q({\bf m}) = 0$ which potentially
cause the modes to pair up and form a gap,
leaving no residual gapless modes in
the low-energy theory.
Let us consider a single quantum Hall state where
we arbitrarily choose there to be an edge
in the bulk of the state.  The edge state will
have the same $K$ matrix structure as the edge
between any two states that have the
same topological structure.  
Such a fictitious edge in the bulk 
cannot support gapless charged
excitations, as is consistent with our above analysis.
Generically, we then expect that at the edge between
two states with the same topological structure there
should be no residual gapless modes.

We can also consider the edge between two states
with different topological structure at
the same filling fraction,
as is appropriate at the
edge between the pseudospin singlet state or
the particle-hole conjugate state and the 
state with independent $\nu=1/3$ Laughlin 
states in each layer.
Generically there is only
one $K({\bf m})=0$ tunneling operator at
the edge between double component states
of different topological order.
It isn't possible for two sets of edges to
pair up and form a gap, due to the mismatch in
topological structure.  There will always be
a pair of residual {\it neutral} gapless edge modes 
left over in the low-energy 
theory, with a gap for making charged
excitations.
As an example we shall consider the edge 
between two states at $\nu =2/3$ possessing
different topological order.
This will be based
on the $\{ {\bf K},{\bf q} \}$ pair
\begin{equation}
{\bf K} = \left(
\begin{array} {cccc}
1 & 2 & 0 & 0 \\
2 & 1 & 0 & 0\\
0 & 0 & -3 & 0 \\
0 & 0 & 0 & -3\\
\end{array}
\right)
\ \ \
{\bf q} = \left(
\begin{array} {c}
1 \\
1 \\
1 \\
1 \\
\end{array}
\right)
\label{edge2}
\end{equation}
The operator $\Psi_{\bf m}$
with
${\bf m} = \{1,1,-1,-1 \}$ has $K({\bf m}) = 0$
and $q({\bf m})=0$,
thereby allowing two charged modes to pair off
and form a gap, leaving two neutral
gapless modes in the low-energy effective theory.
The mismatch between two distinct
quantum Hall states at the same filling fraction
with different topological order implies
the existence of residual neutral gapless excitations
at the boundary.

In considering the edge between two distinct quantum
Hall states at the same filling fraction, we
expect two possible scenarios.  If the two states
possess the same topological structure, such
as the pseudospin singlet state and
the particle-hole conjugate state, their
respective edges will pair up and form gaps, leaving
behind no residual gapless states. If the two
states have different topological structure,
only one set of modes will split off and form 
a gap.  Two neutral, gapless modes will remain
in the low-energy theory, a residual
side effect of the mismatch in topological
order.  

\section{Phase Diagram from Finite Size Studies}
The finite size studies that we report
were carried out using periodic
boundary conditions (see Appendix)
and delta function wavefunctions
to represent the two layers in the
double layer model, with the
electrons confined to the 
lowest Landau level as is standard
practice. 
Our studies were performed at filling fraction
$\nu = N_e / N_{\phi} = 2/3$ with
six electrons and nine flux quanta.
In the following,
length will be measured in units
of $l$, the magnetic length $\sqrt{\hbar / e B}$,
and energy in units of $e^2 / 4 \pi e l.$
The system of electrons interacting
via the Coulomb interaction is exactly diagonalized
numerically, with the spectrum 
as a function of ${\bf k}$ (see Appendix)
providing the
fundamental information on the system.
We have diagonalized the system with the 
Hamiltonian given by
\begin{eqnarray}
H &=& -{\Delta_{sas} \over 2} \sum_i (c^{\dagger}_{i,1}
c_{i,2} + h.c.) \nonumber\\
&+& {1 \over 2} \sum_{i,j,k,l} \langle i \alpha_i,
j \alpha_j | V | k \alpha_k,l \alpha_l \rangle 
c^{\dagger}_{i \alpha_i} c^{\dagger}_{j \alpha_j}
c_{k \alpha_k} c_{l \alpha_l}
\label{ham}
\end{eqnarray}
where
\begin{equation}
V(r) = {1 \over 
4 \pi \epsilon}
{e^2 \over (r^2 + d^2 (1-\delta_{\alpha_i, \alpha_j}))^{1/2}}
\label{inter}
\end{equation}
and
$\alpha_i$ is the layer index which denotes 
in which
of the two layers the electron resides.

All
calculations were performed 
using square boundary conditions where
$\theta= \pi /2$ and $|{\bf L}_1| = |{\bf L}_2|.$
It is known~\cite{Hald5} that incompressible
states are remarkably insensitive 
to the particular boundary conditions
chosen, as long as the shortest
length scale of the geometry is
larger than average interparticle
spacing.
While the exact details of the 
excitation spectrum in the system
under investigation 
will depend on our choice of
conditions, 
the qualitative conclusions 
concerning the incompressible
ground states should not.

In the following we shall use the term
pseudospin to refer to the 
subband layer index, with the
the up-spin corresponding to the symmetric
combination of layer states and the
down-spin the antisymmetric. 
Tunneling acts as a Zeeman term in
the Hamiltonian which
tends aligns the electron pseudospin in
the up state, or the
symmetric combination of layer states.
We will vary both the tunneling, denoted
in the Hamiltonian by $\Delta_{sas}$,
and the distance between the double layer 
planes, denoted by $d/l$, investigating both
the ground state and the dependence of the
excited states on these parameters.

\subsection{Spin-Singlet State}
It was realized some time ago~\cite{Halp1}
that when the electron correlations
are of the same order as the Zeeman energy 
associated with the spin states,
it is important to consider the spin degrees
of freedom in constructing the ground state.
Previous numerical and experimental 
studies~\cite{Eisen2,Mak1} on the
$\nu =2/3$ system reveal that the ground state
of the non spin-polarized system, in the limit
of vanishing Zeeman energy, is plausibly a spin-singlet.
If we consider a phase diagram where we vary both
the interlayer separation $d/l$ and the interlayer
tunneling $\Delta_{sas}$, along the
line $d=0$ 
the Hamiltonian is invariant under 
pseudospin rotations, allowing
a direct mapping between 
pseudospin and electron spin
in the presence of Zeeman term.
We know from our previous edge analysis~(\ref{alg})
that the $K$ matrix of the effective theory of a
spin-singlet
ground state must be of the form
\begin{equation}
{\bf K} = \left(
\begin{array} {cc}
m \pm 1 & m \\
m & m \pm 1 \\
\end{array}
\right)
\ \ \
{\bf q} = \left(
\begin{array} {cc}
1 \\
1 \\
\end{array}
\right)
\label{sform}
\end{equation}
where $m$ is even.
Jain~\cite{Jain} has proposed a wavefunction to
describe the spin-singlet state at $\nu = 2/3$
based upon the effective theory where
$\{ {\bf K},{\bf q} \}$ is given by
\begin{equation}
{\bf K} = \left(
\begin{array} {cc}
1 & 2 \\
2 & 1 \\
\end{array}
\right)
\ \ \
{\bf q} = \left(
\begin{array} {cc}
1 \\
1 \\
\end{array}
\right).
\label{kmats}
\end{equation}
We cannot strictly interpret
this $K$ matrix in terms of the 
Halperin-Laughlin 
wavefunction, as thermodynamic
stability requires the $K$ matrix
for such a wavefunction to be positive 
definite, so we must
interpret the wavefunction
in the planar geometry to be 
\begin{eqnarray}
\Psi({z_i, \sigma_i}) &=& 
\prod_{i<j} ({\partial_{z_i}}
- {\partial_{z_j}})^{\delta_{\sigma_i,\sigma_j}}
\prod_{i<j} (z_i - z_j)^2 \nonumber\\
&& \times \prod_{i<j} {\rm exp}
\biggl({i{\pi \over 2} {\rm sgn}
(\sigma_i - \sigma_j)}\biggr) \nonumber\\
&& \times \prod_{i}
{\rm exp} \biggl(-{1 \over 4 l^2} {\vert z_i \vert}^2
\biggr)
\end{eqnarray}
where we have denoted 
$\partial_{z_i} = {\partial \over \partial {z_i}}.$ 
This wavefunction is a spin-singlet at the
correct filling fraction with the
correct topological properties.

In Fig.~[\ref{one}] we calculate the overlap of the
exact ground state with Coulomb interactions
with Jain's proposed
spin-singlet wavefunction as we vary the
Hamiltonian parameters $\Delta_{sas}$ and
$d/l$ (note that $\Delta_{sas}$ and
tunneling are used interchangably).  
Several points should be noted.  
First, the system is very sensitive to the
effects of tunneling, or equivalently,
a magnetic field in pseudospin space.  The
overlap with the spin-singlet state rapidly
falls off with the introduction of even a slight
amount of tunneling.  This will tend to make the 
experimental observation
of the spin-singlet
state in multilayer samples extremely difficult.
Second, the system is reasonably robust against
a separation between the two layers, falling 
to zero at $d \approx 1.1 l.$ 
Thus we find significant overlap between the
exact ground state and the proposed spin-singlet
even in regions where the Hamiltonian no 
longer commutes with the pseudospin algebra.  The
spin-singlet character of the state is a
manifestation of the
electron correlations rather than the underlying
Hamiltonian.

While we expect that the numerical data
gives us qualitative data on the transitions
discussed, it should be noted that 
finite size effects will influence the
exact positioning of the transitions in the $d/l - \Delta_{sas}$
plane in relation to the thermodynamic limit, and
should therefore be taken with caution.

\subsection{Particle Hole Conjugate of $\nu =1/3$ Laughlin State}
The particle-hole conjugate of the
standard Laughlin $\nu=1/3$ gives us
an incompressible liquid at filling fraction
$\nu=2/3$ which
is denoted in the 
effective theory by the same
$K$ matrix as that of the spin-singlet 
state~(\ref{kmats}).
The matrix nature of the effective theory
does not reflect the correlations between
electrons of opposite pseudospin, as the 
particle-hole conjugate state
is pseudospin polarized, but rather reflects its
composite nature.
While 
to our knowledge no simple wavefunction 
has been proposed for
the exact particle-hole conjugate state,
a trial hierarchy wavefunction which very
effectively captures the electron correlations
has been developed~\cite{Wen2}.  
We may write the hierarchy
$\nu =2/3$ state as
\begin{eqnarray}
\Psi(\{ z_i,\sigma_i \}) &=&
\prod_{i^{\prime}} \int d \omega_{i^{\prime}}
d \omega^{*}_{i^{\prime}} 
\prod_{i^{\prime}<j^{\prime}}
(\omega^{*}_{i^{\prime}}
- \omega^{*}_{j^{\prime}})^2 \ {\vert \omega_{i^{\prime}}
- \omega_{j^{\prime}} \vert}^{2} \nonumber\\
&&\times \prod_{i,j^{\prime}}
(z_i - w_j^{\prime}) 
\prod_{i<j} (z_i - z_j) \nonumber\\
&&\times \prod_{i^{\prime}} 
{\rm exp} \biggl(-{1 \over 2 l^2}
{\vert \omega_{i^{\prime}} \vert}^2 \biggr) \nonumber\\
&&\times \prod_{i}
{\rm exp} \biggl(-{1 \over 4 l^2}
{\vert z_i \vert}^2 \biggr) 
\end{eqnarray}
where $i^{\prime}=1,...,N_h$ and $i=1,...,N_e$
where the number of holes $N_h = N_e/2$
and $N_e$ is the number of electrons.
This hierarchical state is represented by
the effective theory given by
\begin{equation}
{\bf K} = \left(
\begin{array} {cc}
1 & 1 \\
1 & -2 \\
\end{array}
\right)
\ \ \
{\bf q} = \left(
\begin{array} {cc}
1 \\
0 \\
\end{array}
\right)
\end{equation}
which is equivalent to~(\ref{kmats})
upto a similarity transformation.
This effective theory also possesses an
$SU(2)$ invariance.  In this case, the 
two pseudospin variables refer to the 
lowest two Landau levels in the composite
fermion construction.
Within the composite fermion approach,
to obtain the single layer particle-hole
conjugate state one starts with a spin
polarized $\nu = 2$ state, a gapped
system, and adiabatically 
attaches two flux quanta to each electron
opposite to the direction of
the magnetic field that generated the 
$\nu = 2$ state.  As the addition of
two flux quanta doesn't affect the statistics,
our composites are still fermions.  In a mean-field
sense, we start with a pseudospin polarized
state at $\nu=2$
and decrease the $B$ field by two flux
tubes per electron, arriving at a 
pseudospin polarized state at $\nu=2/3$,
which we identify as the particle-hole
conjugate state.  We still have residual gauge
fluctuations associated with the added flux, 
but they shouldn't qualitatively change the physics
as we started with a gapped system.  Therefore
the $\nu =2/3$ polarized state can be identified 
with a polarized $\nu = 2$ integer quantum Hall state.
For the spin-singlet state, out starting
point is a spin-unpolarized $\nu =2$ state
with the first Landau level being filled
for both the up and down spins.  We then
perform the same flux addition process
as we did for the polarized state 
to arrive
at a spin-singlet $\nu =2/3$ fermion state.
Within this approach, which state will
be realized depends on the
ratio of the
effective cyclotron energy
to the effective Zeeman energy of the
composite fermions. 
 
In Fig.~[\ref{two}] we calculate the overlap
of the particle-hole conjugate state with 
the exact ground state as a function of
$\Delta_{sas}$ and $d/l$ ($\Delta_{sas}$
and tunneling being used interchangably
as in Fig.~[\ref{one}]).
The effect of the tunneling in the spin 
analogy can be seen to be a turning on of
a Zeeman energy term in the $\hat z$ direction.
As this energy is increased, eventually all
the spins will
align themselves along the $\hat z$ direction,
resulting in a spin-polarized Laughlin state
with all the electrons occupying a symmetric
combination of the layer indices.
As the Zeeman term is
turned on, the
system abruptly reaches a transition point
where the the overlap of the ground state 
with the spin-singlet state 
falls to zero (Fig.~[\ref{one}]), 
while the overlap with the Laughlin $2/3$
particle-hole conjugate state 
jumps to close to unity.

As the spin-singlet state and the particle-hole
conjugate state are described by the same 
pair $\{ {\bf K},{\bf q} \}$,
they are described by the same effective theory
on the plane and possess the same edge state structure.
While the effective theory approach captures
many of the long-distance properties of
the fluid, it doesn't classify the
spin of the state, which is determined 
by the energy associated with the spin
degree of freedom.
As we turn on the tunneling $\Delta_{sas},$
we find a transition from a region 
where it is energetically
favorable to put the composite fermions
in the first two spin states in the lowest
Landau level to a region where
it is favorable to put them in the first
two Landau levels.  In both cases
we have an $SU(2)$ symmetry, in the one case
between pseudospin states and in the
other between Landau levels.
It is important to notice that
within the composite fermion approach 
the effective cyclotron energy is heavily
renormalized, as small variation in the
effective field essentially makes it 
energetically favorable to place the 
electrons in two pseudospin-polarized 
Landau levels, costing cyclotron energy 
but saving on Zeeman energy.
The composite fermion process of
attaching flux has the effect
of enhancing the ratio
of the effective energy associated with
a pseudospin flip 
to the effective 
cyclotron energy.

\subsection{$(3,3,0)$ Double Layer State}
At non-zero $d/l$, as the Hamiltonian no
longer commutes with psuedospin rotations
the ground state need not be an eigenstate of
pseudospin nor need it possess an $SU(2)$
symmetry as the particle-hole conjugate
state does.
A proposed effective theory for the
ground state at large layer separation $d/l$
is given by
\begin{equation}
{\bf K} = \left(
\begin{array} {cc}
3 & 0 \\
0 & 3 \\
\end{array}
\right)
\ \ \
{\bf q} = \left(
\begin{array} {cc}
1 \\
1 \\
\end{array}
\right)
\end{equation}
which has a direct interpretation
as a $(3,3,0)$ Halperin-Laughlin multi-component
wavefunction representing independent $\nu=1/3$
Laughlin states in each layer.
In Fig.~[\ref{three}] we calculate the overlap of the exact
ground state with the $(3,3,0)$ state as a function
of $\Delta_{sas}$ and $d/l$ (note again that
tunneling and
$\Delta_{sas}$ are being used 
interchangably).
The transition between the $(3,3,0)$ state and
the particle-hole conjugate state (Fig.~[\ref{two}])
represents a competition between minimizing the 
coulomb energy between the electrons and the tunneling
energy cost of localizing the electrons on independent
planes.
For any given tunneling $\Delta_{sas}$
it is possible to find a large enough $d/l$ such that it is
energetically favorable to form the independent
$\nu =1/3$ states in each layer, costing in tunneling
energy but gaining even more in minimizing the electrostatic
ground state energy.
Equivalently, given any interlayer distance we can
find a tunneling energy which will outweigh the
favorable interaction energy associated with having
the electrons as far apart as possible.  

\subsection{Phase Diagram}
We can identify three separate phases for
the $\nu=2/3$ system as a function of
tunneling $\Delta_{sas}$ and the
layer separation $d/l$. 
We calculate the phase diagram for the
$\nu =2/3$ system in Fig.~[\ref{four}]
from the overlap data
by matching a system with a particular phase if 
the
overlap of the ground state
with the characteristic state 
is greater than
some cutoff, chosen to be .75 in Fig.~[\ref{four}].
One important feature to note is the existence of a 
triple point in our numerical studies at
$d \approx 1.1l$ and $\Delta_{sas} \approx .01$
where all three phases 
will be in coexistence.
It is difficult to extract the experimental
parameters where such a triple point might occur
from our finite size numerical data, as such
quantitative information will be sensitive to finite
size effects such as geometry and particle 
number.  
We conjecture that the triple point 
where all
three discussed phases will be in coexistence
will persist in the thermodynamic
limit, in the neighborhood of the 
physical parameters
suggested by our studies.
The observation of such a triple point remains
an interesting experimental possibility.

\section{Finite Size Study of Transitions}
Whenever a system exhibits different macroscopic
phases as a function of system parameters, it
is natural to ask questions about the
transitions between such phases.  While
finite size studies are unable to address questions
about
the thermodynamic features of such
transitions, they are able
to shed light on qualitative changes in
structure that a bulk liquid may undergo 
in going from one phase to another.
In order to investigate the transitions
involved in our double layer $\nu =2/3$ system,
we have exactly diagonalized our finite
system using periodic boundary conditions
as a function of the system parameters
$\Delta_{sas}$ and $d/l$.
We can follow the development
of individual energy levels of the system
as these
parameters are varied by identifying  
the quantum numbers of the state in question
such
as parity under reflection and
the translational quantum number ${\bf k}$
and noting that under adiabatic perturbation
the energy levels should be smooth and
connected in our finite system.
In the following figures we plot the
energy levels of the system as a function
of either $\Delta_{sas}$ and $d/l$, at each
step subtracting off the average energy of the
system in order eliminate background
energies of the system.  The energy levels
which are relevant to each transition are connected
for emphasis
during their development. 

When using periodic boundary conditions it is
important to note that 
each eigenstate has a generic degeneracy
associated with center-of-mass translations
given by $q$ if the filling fraction
is $\nu = p/q$.
The formalism used to classify 
states (see Appendix) extracts this degeneracy  
explicitly.  Since we are always working at
fixed filling fraction, this
degeneracy will be unimportant.

\subsection{Variation of $d/l$ at $\Delta_{sas}=0$}
In Fig.~[\ref{five}], the variation of the energy levels
as a function of $d/l$ for $\Delta_{sas}=0$ is shown,
with the average energy at each $d/l$ subtracted off.
The transition being witnessed
is from the spin-singlet phase to the
$(3,3,0)$ phase as $d/l$ is increased.
At small values of $d/l$ the ground state
is well represented by Jain's spin-singlet
state, with a well-defined energy gap to all excitations
indicative of an incompressible phase.

At large values of $d/l$ the ground state is
given by a three-fold degenerate multiplet
of states, each being a ${\bf k}=0$ eigenstate
which in the thermodynamic limit
becomes rotationally invariant.
There is again a well-defined energy gap between
the ground state multiplet and all excited states.
There is a slight splitting of the degeneracy
due to finite size
effects but this feature will disappear as the
size of the system is increased.

The key feature to note is the change in ground state
degeneracy in going from the spin-singlet state
to the $(3,3,0)$ state.  Such a change
is to be expected from our
effective theory considerations.
As noted previously the degeneracy of a state given by
the effective theory with the pair $\{ {\bf K},{\bf q} \}$
on a torus is given by $\vert {\rm Det} K \vert$.
The spin-singlet state must then
have an overall degeneracy of three while the
$(3,3,0)$ state has a degeneracy of nine.
As mentioned above our formalism extracts a three-fold
center-of-mass degeneracy generic to
states at $\nu =2/3$, leaving us
with a residual three-fold degeneracy for the
$(3,3,0)$ state and a non-degenerate spin-singlet
state, consistent with our numerical data.

Another point to be noted is that 
one of the energy levels coming down in the $(3,3,0)$
state triplet has the same symmetry as the spin-singlet
state, resulting in an energy level repulsion
as they cross.  As the transition point approaches,
the triplet of states comes down, crossing with
the spin-singlet state.

It is interesting to note that even in considering
bulk transitions, as we are in our finite size studies,
the systems display a residual side effect from the
topological mismatch between the two states in
coexistence.  In our edge state analysis we found that
at the edge between the spin-singlet and
$(3,3,0)$ state there will be residual neutral
gapless modes due to the difference in topological 
structure, while in our bulk finite size studies we
find a change in ground state degeneracy.  Both
features are generic and stable against 
perturbation, indicative of
the true topological character of the 
incompressible fluids.

\subsection{Variation of $\Delta_{sas}$ at $d/l=0$}
In Fig.~[\ref{six}] we investigate the energy level structure 
by fixing $d/l=0$ and
varying $\Delta_{sas}.$
We are witnessing the transition from the
spin-singlet ground state to the particle-hole
conjugate state as a function of the Zeeman
energy associated with the pseudospin degree
of freedom.
The system quickly undergoes a transition from the
spin-singlet state to the particle-hole
conjugate state indicated by a simple level crossing.
The spin-singlet state is extremely sensitive to the
effects of the effective magnetic field, rapidly finding
it energetically favorable to place the composite
fermions in the second Landau level rather than the
first spin-reversed Landau level.

As the two phases possess the same topological structure
they also possess the ground state degeneracy.  They do
however have different pseudospin symmetries, allowing the
energy levels to cross without repulsion.  
Such an energy level crossing will become a first-order
transition in the thermodynamic limit.

\subsection{Variation of $\Delta_{sas}$ at $d=2.0l$}
In Fig.~[\ref{seven}] we investigate the energy structure
at $d = 2.0l$ as we vary $\Delta_{sas}.$  
We are seeing the transition from
the $(3,3,0)$ state to the particle-hole conjugate 
state as we turn up the tunneling.
The transition is qualitatively the same as
the spin-singlet to $(3,3,0)$ transition, with
a degeneracy transition due to the topological
mismatch between the two states.
Again we see a level crossing driven by
variation in the sample parameters, where
two states involved in the crossing mix
and cause energy level repulsion.
This transition is particularly
relevant as 
experimental evidence supporting 
such a phase transition
already exists~\cite{Suen}.
The experiments of Suen, et.al. were
performed
using a single wide quantum well geometry,
making quantitative comparison with our
idealized double layer calculations 
difficult~\cite{Ian3}.

\section{Nature of the Transitions}
We believe that the transitions seen in our
finite size studies represent 
first-order phase transitions in the
thermodynamic limit at low
temperatures in a clean system.
Traditionally, one uses broken symmetries
and their associated order parameters to
classify and organize many-body systems,
often allowing the construction of
low energy effective theories 
based on these order parameters which 
capture the essence of the correlated 
states as well as predicting effects that
are not accessible in a microscopic
approach.
The effective theories employed in the 
description of the fractional quantum Hall
effect are not based on an order parameter
derived from a broken symmetry.  Rather,
these effective theories embody a new
type of order called topological order
which is stable against perturbation
and manifests itself in such 
properties as the ground state degeneracy
when the system is defined on a 
topologically non-trivial closed space.
We therefore do not consider the fractional
quantum Hall states to be broken symmetry 
states.

As the fractional quantum Hall states are not
broken symmetry states, they cannot undergo the
usual second order phase transition 
where the 
order parameter smoothly goes to zero
in the vicinity of a transition point,
and is zero on the other side of the
transition.  
It would appear that due to the
fact that the order embodied by
the fractional quantum Hall states
is discrete and topological in origin
that it would be impossible to smoothly go
from one state to another 
with different topological order
as a function of system
parameters in a clean system.
Let us consider a point in the
$d/l - \Delta_{sas}$ plane where either
the $\nu =2/3$ spin-singlet state
or the particle-hole conjugate
state is in coexistence with the $(3,3,0)$
state.
From our edge state
analysis we know that when two phases
having different topological order are in
coexistence 
there will be at least one pair of residual 
neutral gapless modes 
at the boundary between phases.
The finite energy of these modes will
localize the boundary between the
two phases, forcing the transition
to occur by nucleation, resulting in
a first-order phase transition.
If two states have different topological order,
there is no way to smoothly go from one
state to another due to the discrete
nature of topological order.

In the transition between two states
that possess the same topological order
there is no such mismatch.  
If we consider the interface between
two such states,
such as the spin-singlet state and the
particle-hole conjugate state, we
expect that the two pairs of edge
modes will pair up and form a gap, 
leaving no residual gapless modes.
In a clean system the two states
have different pseudospin
symmetries, as the particle-hole conjugate state
is spin-polarized while the spin-singlet state
is a true pseudospin singlet.
This difference is reflected in the
flux-number shift of the two states on
the sphere: ${\cal S}=0$ for the particle-hole
conjugate state and ${\cal S}=1$ for the spin-singlet.
The transition can then go by a simple level 
crossing, resulting in a first-order transition
in the thermodynamic limit.

There exists the possibility that there
might be some intermediate state that
exists between the two principal states
undergoing the transition.
This case is really a two step process rather
than a direct transition: Hall state to
intermediate state, then intermediate state to
Hall state.  In principal the intermediate state
could be incompressible, but we can then apply
the same arguments used above to show
that the two principal
states undergoing the transition cannot be smoothly 
connected.
While this scenario
is an experimental possibility,
in the $\nu=2/3$ system no evidence exists for this
type of intermediate
transition.  As such, we
conjecture that all the transitions 
involved in the $\nu=2/3$ system will be first-order.

\section{Conclusions}
In conclusion, we have examined the 
structure of the
phase diagram of the $\nu =2/3$
double layer electron system
as a function of $d/l,$
the distance between the layers, and
$\Delta_{sas},$ the tunneling 
parameter for the system.
A phase diagram consisting of three
different phases, each belonging
to a different, distinct universality
class, was calculated.  A triple 
point is conjectured where all 
three phases are stable.
A gapless, neutral Luttinger liquid
structure is predicted at 
the interface between either the spin-singlet state
or the particle-hole conjugate state and
the $(3,3,0)$ state.  At the interface
between the spin-singlet state and the
particle-hole conjugate state no residual gapless
modes are expected.
It is conjectured that there
should be first-order transitions between all three
phases, indicated by distinct level
crossings in the finite system 
energy levels.

\acknowledgements
We wish to acknowledge valuable discussions 
with M. Shayegan, M. Cole, and C. Nayak.
I. A. M. would like to acknowledge
financial support from NSF grants
DMR-90-22681 and DMR-92-24077.

\appendix
\section*{Periodic Boundary Conditions}
We wish to impose generalized boundary conditions
by requiring that all physical quantities
be invariant under translation of
any particle by the set of
translations
${\bf L}_{mn}= m {\bf L}_1 + n {\bf L}_2$
such that
\begin{equation}
\vert {\bf L}_1 \times {\bf L}_2 \vert = 2 \pi N_{\phi} l^2
\label{fluxnu}
\end{equation}
where
$N_{\phi}$ is the number of flux quanta.
We impose the general boundary conditions on
the wavefunction for any particle $i$
\begin{equation}
t_i({\bf L}_{mn}) \vert \Psi \rangle
= (\eta_{mn})^{N_{\phi}} e^{i {\bf \Phi}_0 \cdot
{\bf L}_{mn}} \vert \Psi \rangle
\label{pbc}
\end{equation}
where
$\eta_{mn} = (-1)^{(m+n+mn)}$
and we will choose
${\bf \Phi}_0 = 0$ as our boundary condition.
The operators $t({\bf a})$ are the translation operators 
in the presence of a magnetic field~\cite{Hald4}
which obey the non-commutative
algebra
\begin{equation}
t({\bf a}) t({\bf a}^{\prime}) = t({\bf a} + {\bf a}^{\prime})
e^{i {{\bf a} \times {\bf a}^{\prime} \over 2 l^2}}.
\end{equation}
We shall denote the two dimensional
coordinate as $a = a_x + i a_y$
and ${\bar a} = a_x - i a_y.$

In the periodic geometry, Halperin's extension of
Laughlin's wavefunction, suitable for double layer
systems, can be written
\begin{equation}
\Psi^{(m_1, m_2, n)} [z_i, z_i^{\prime}] =
\Psi_{c.m.} [Z, Z^{\prime}] 
\ \Psi^{(m_1, m_2, n)}_{\rm rel} [z_i, z_i^{\prime}]
\label{wave}
\end{equation}
where 
\begin{eqnarray}
\Psi^{(m_1, m_2, n)}_{\rm rel} [z_i, z_i^{\prime}] &=&
\prod_{i<j} [\phi(z_i - z_j)]^{m_1} 
\prod_{i<j} [\phi(z_i^{\prime} - z_j^{\prime})]^{m_2} \nonumber\\ 
& & \times \prod_{i,j} [\phi(z_i - z_j^{\prime})]^n
\label{rel}
\end{eqnarray}
$Z = \sum_i z_i,$ $Z^{\prime} = \sum_i z^{\prime}$
and the unprimed coordinates refer to
electrons in the first layer
while the primed refer to electrons
in the second.
This wavefunction is denoted 
$(m_1,m_2,n).$
The basic building block that we have built our wavefunction
from is the 
quasiperiodic function $\phi(z)$ which can be
written
\begin{equation}
\phi(z) = w(z) {\rm exp} 
\biggl[-\bigl({z^* z \over 4 N_{\phi} l^2}
\bigr) \biggr].
\label{parto}
\end{equation}
where
\begin{equation}
w(z) = {\rm exp} \biggl({z^2 \over 4 N_{\phi} l^2} \biggr)
\Theta_1(\kappa z \vert \tau)
\label{partt}
\end{equation} 
and
$\Theta_1(u|\tau)$ is the odd elliptic theta function,
$L_{mn} = \kappa^{-1} (m + n \tau)$
and
$\tau = {L_2 \over L_1} e^{i \theta}$
where ${\bf L}_1 \cdot {\bf L}_2 = 
\vert {\bf L}_1 \vert \vert {\bf L}_2 \vert {\rm cos}
\theta.$
We have used the symmetric gauge
${\bf A} = {B \over 2} {\bf r} \times {\hat z}$
in expressing the function $\phi.$
We constrain $m_1$ and $m_2$
to be odd for fermi statistics.  
Note that in writing the wavefunction in this
form we have expressed the correlations between the
electrons but for notational simplicity
suppressed the pseudo-spin part of the
wavefunction which would
properly anti-symmetrize the overall wavefunction.
This form of the double layer wavefunctions has been
discussed previously~\cite{Ian2}.

We consider a system of electrons confined to 
two parallel planes subject to periodic
boundary conditions (PBCs) confined to the lowest
Landau level.  The symmetry analysis of this system,
as introduced by Haldane~\cite{Hald4}, allows 
us to construct a Hilbert space which extracts
the center-of-mass degeneracy 
as well as providing a correct
classification of states allowing comparison
with studies performed in other geometries.
We can therefore classify the eigenstates of a
translationally invariant Hamiltonian
obeying $[H,T({\bf a})]=0$
by the quantum number $\bf k$ defined to be
\begin{equation}
T\bigl({{\bf L}_{mn} \over \bar N} \bigr) \vert \Psi \rangle
= (\eta_{mn})^{pq}
{\rm exp}
\biggl( i {{\bf k} \cdot {\bf L}_{mn}
\over \bar N} \biggr )\vert \Psi \rangle
\label{topth}
\end{equation}
where
$N_e = \bar N p$ and $N_{\phi} = \bar N q,$
allowing us to write $\nu = N_e/N_{\phi} = p/q.$
In the
thermodynamic limit the states characterized by
${\bf k}=0$ become rotationally invariant,
implying that the signature of an incompressible
quantum Hall state is a ${\bf k}=0$ ground state
with an energy gap to all excited states.
As the operator~(\ref{topth}) commutes with the
center of mass operator $T({{\bf L}_{mn} \over N_{\phi}})
= \prod_i t_i({{\bf L}_{mn} \over N_{\phi}})$
each eigenstate has a 
$q$-fold degeneracy associated with the
action of the center of mass operator.

\begin{figure}
\caption{Overlap of ground state 
with the spin-singlet state as a 
function of $d/l$ and 
tunneling.  Tunneling is denoted
in the Hamiltonian as $\Delta_{sas}$ and
is measured in units of $e^2 / 4 \pi e l.$}
\label{one}
\end{figure}

\begin{figure}
\caption{Overlap of ground state
with v=2/3 Laughlin particle hole conjugate
state as a function of $d/l$ and 
tunneling.  Tunneling is
denoted in the Hamiltonian as
$\Delta_{sas}$ and is measured
in units of $e^2 / 4 \pi e l.$
Note the change in perspective of this
figure 
from the previous figure.}
\label{two}
\end{figure}

\begin{figure}
\caption{Overlap of the ground state 
with the $(3,3,0)$ state as a function
of $d/l$ and tunneling.
Tunneling is denoted in the 
Hamiltonian as $\Delta_{sas}$
and is measured in units of $e^2 / 4 \pi e l.$
Note the change in perspective of this figure from
the previous two figures.}
\label{three}
\end{figure}

\begin{figure}
\caption{Phase Diagram of $\nu = 2/3$ system 
as a function of $\Delta_{sas}$ and $d/l$.
$\Delta_{sas}$ is measured in units
of $e^2 / 4 \pi e l.$}
\label{four}
\end{figure}

\begin{figure}
\caption{Variation in energy levels
as $d/l$ is varied at $\Delta_{sas}=0.$}
\label{five}
\end{figure}

\begin{figure}
\caption{Variation in energy levels
as $\Delta_{sas}$ is varied at $d/l=0$.
$\Delta_{sas}$ is measured in units
of $e^2 / 4 \pi e l.$}
\label{six}
\end{figure}

\begin{figure}
\caption{Variation in energy levels
as $\Delta_{sas}$ is varied at $d/l=2.0$.
$\Delta_{sas}$ is measured in units
of $e^2 / 4 \pi e l.$}
\label{seven}
\end{figure}

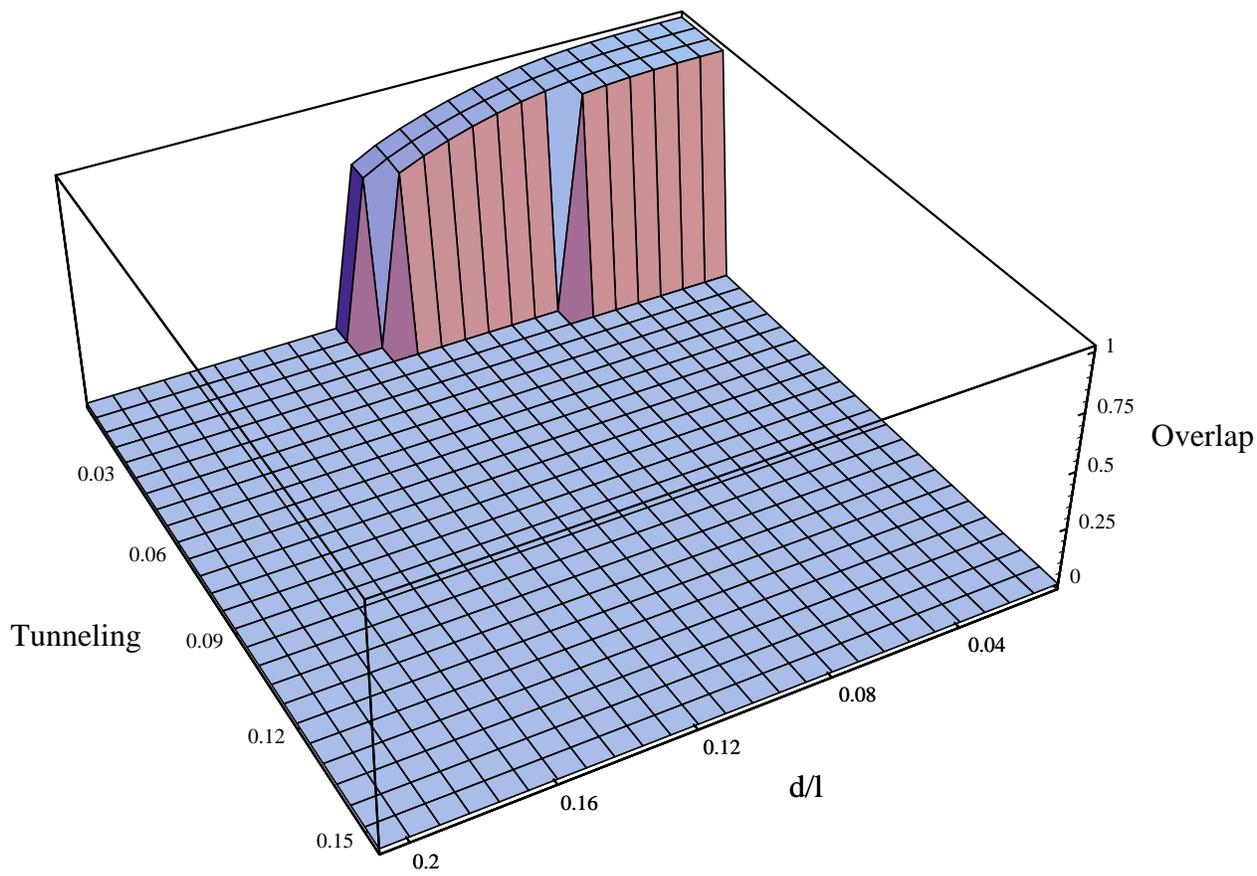
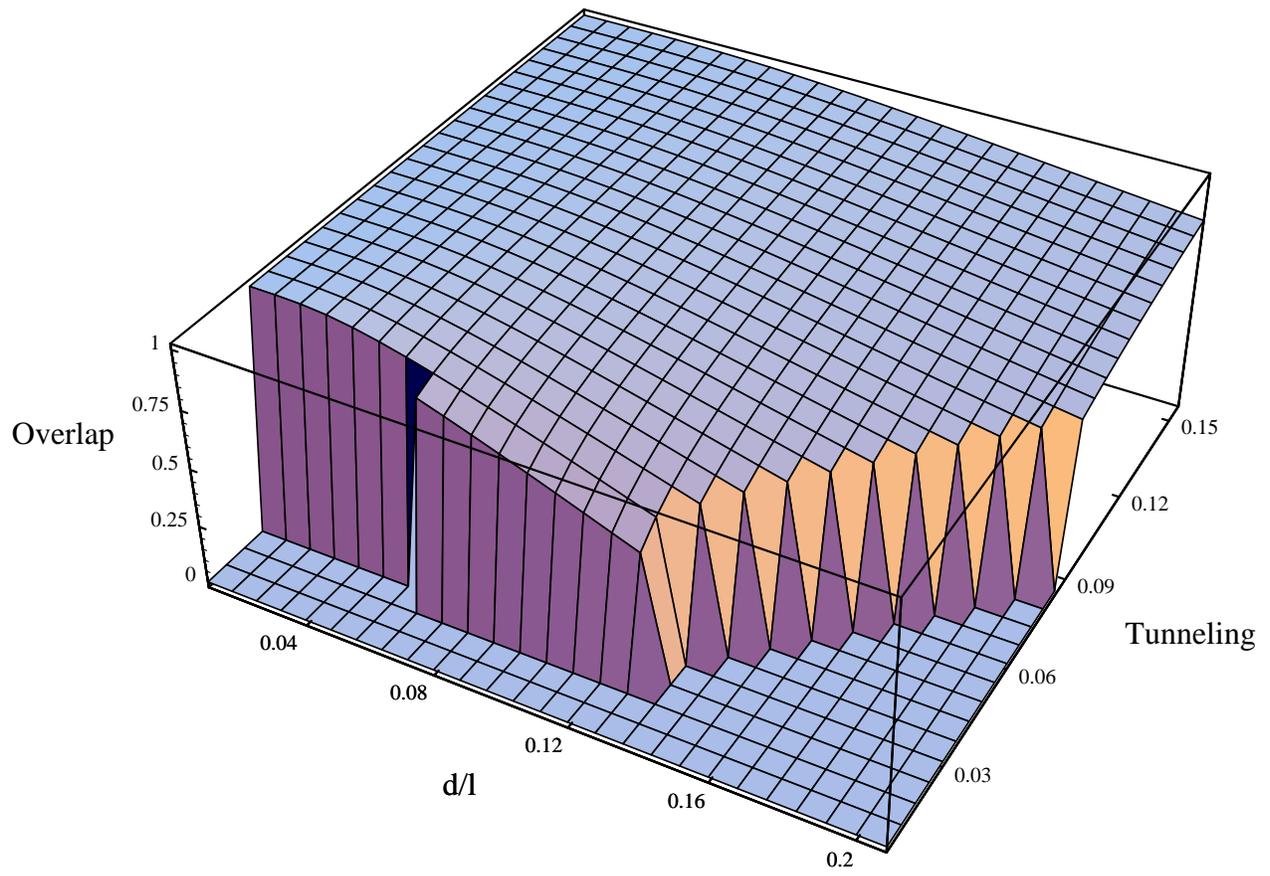
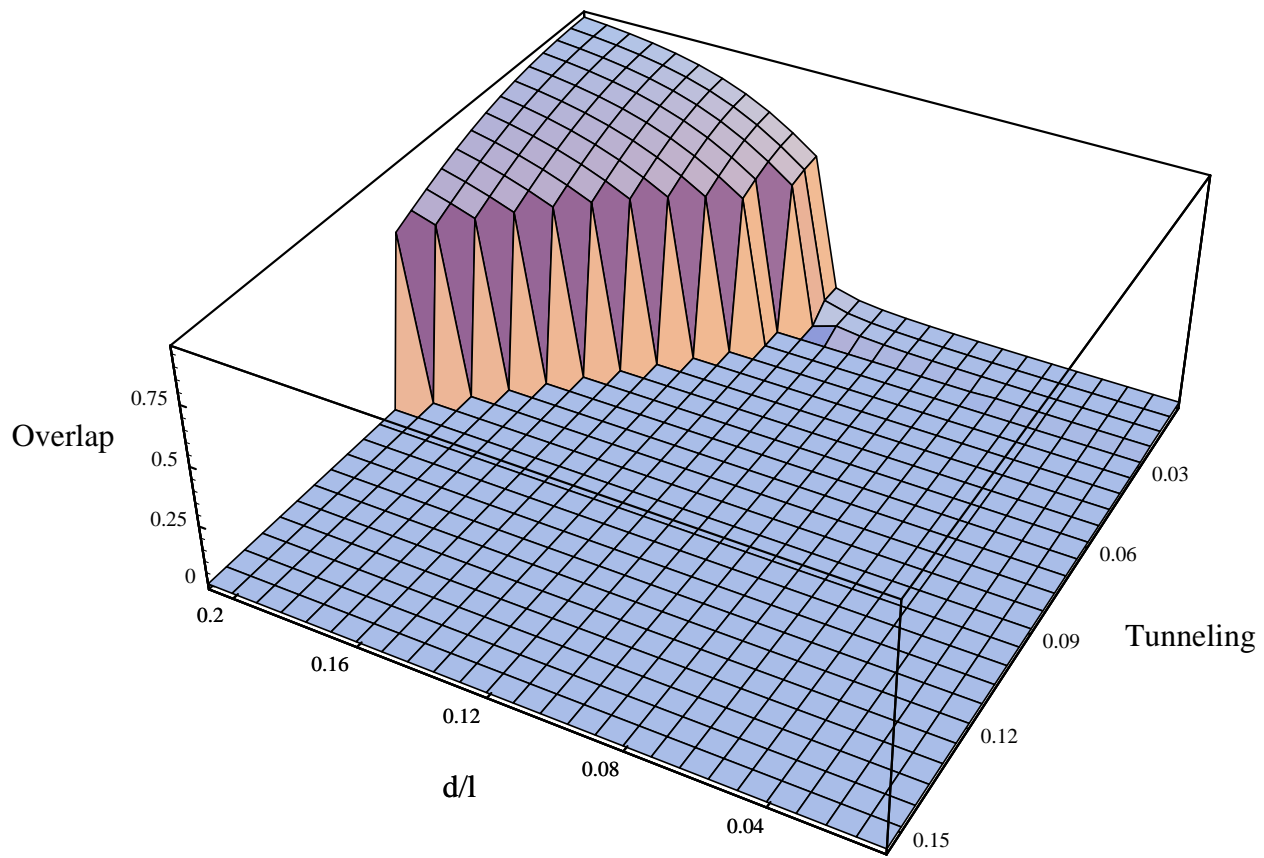
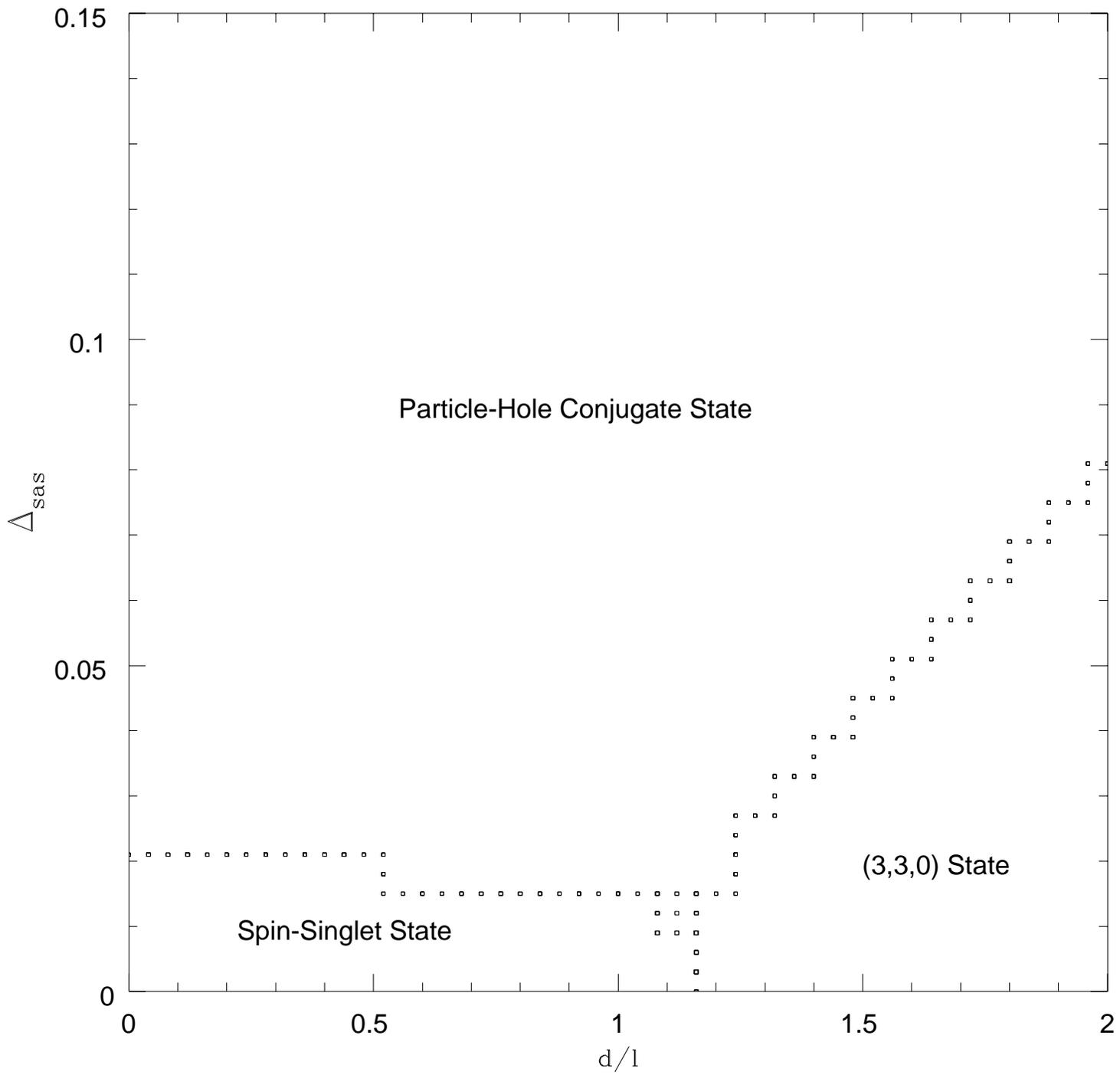
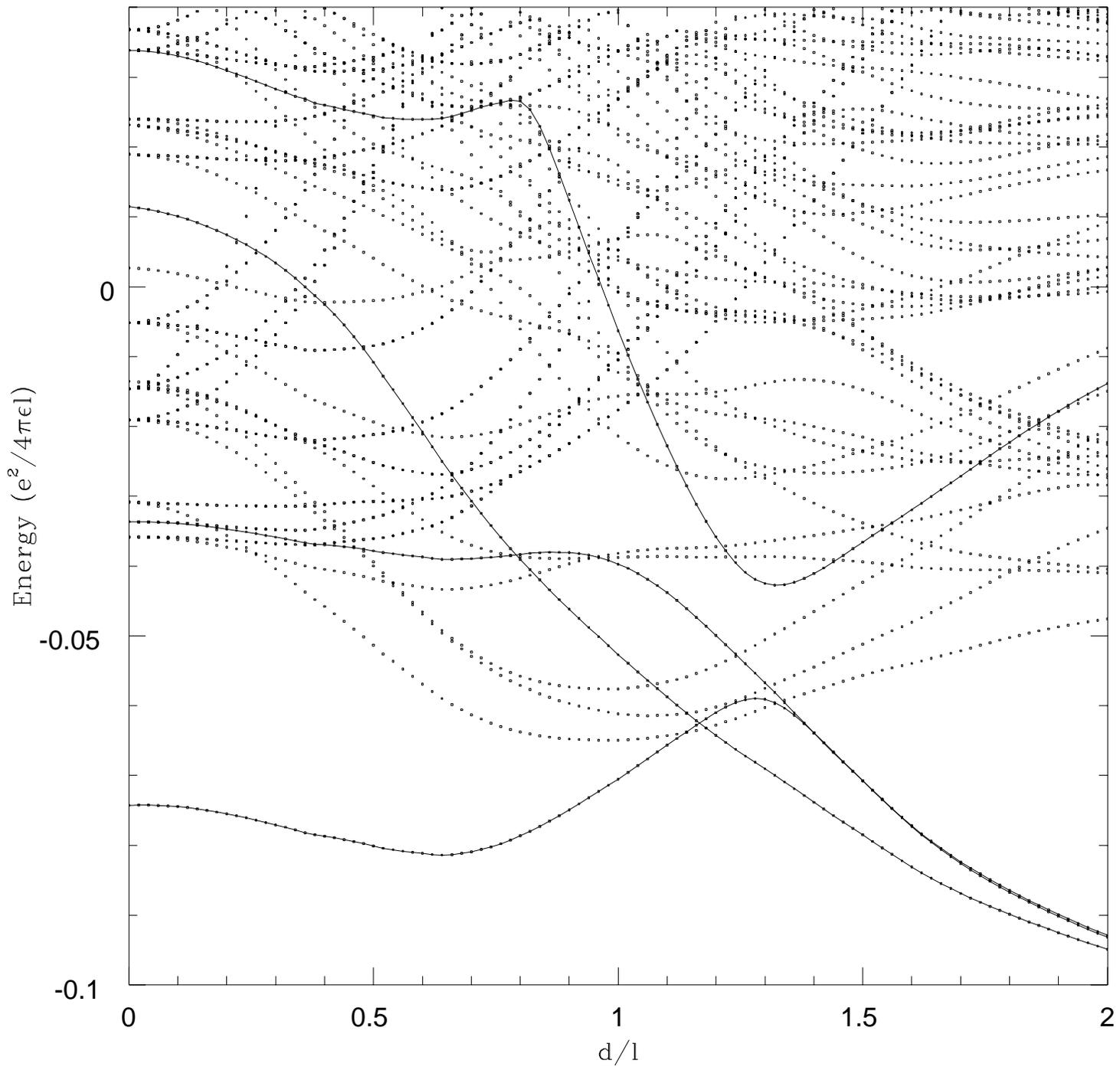
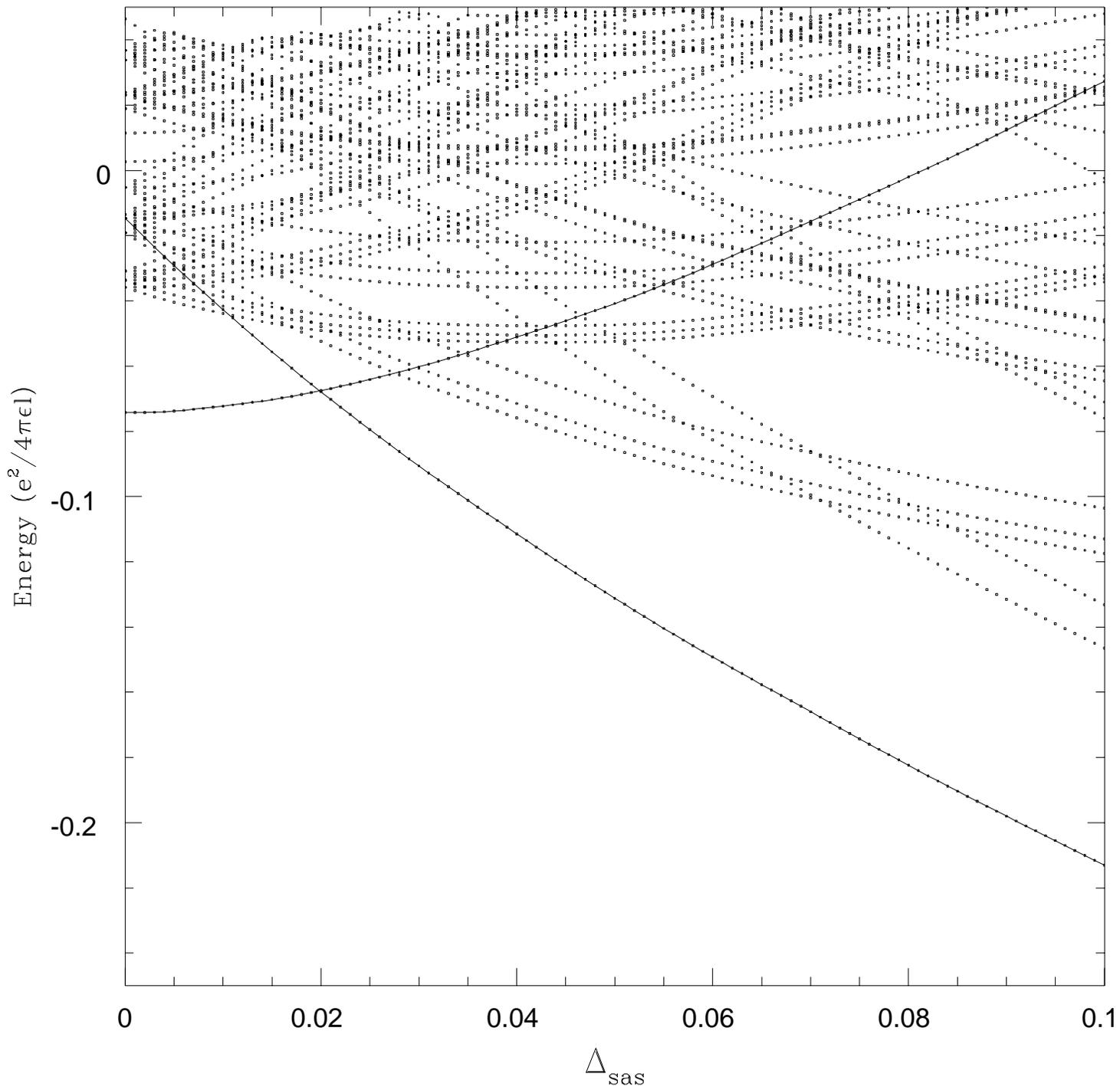
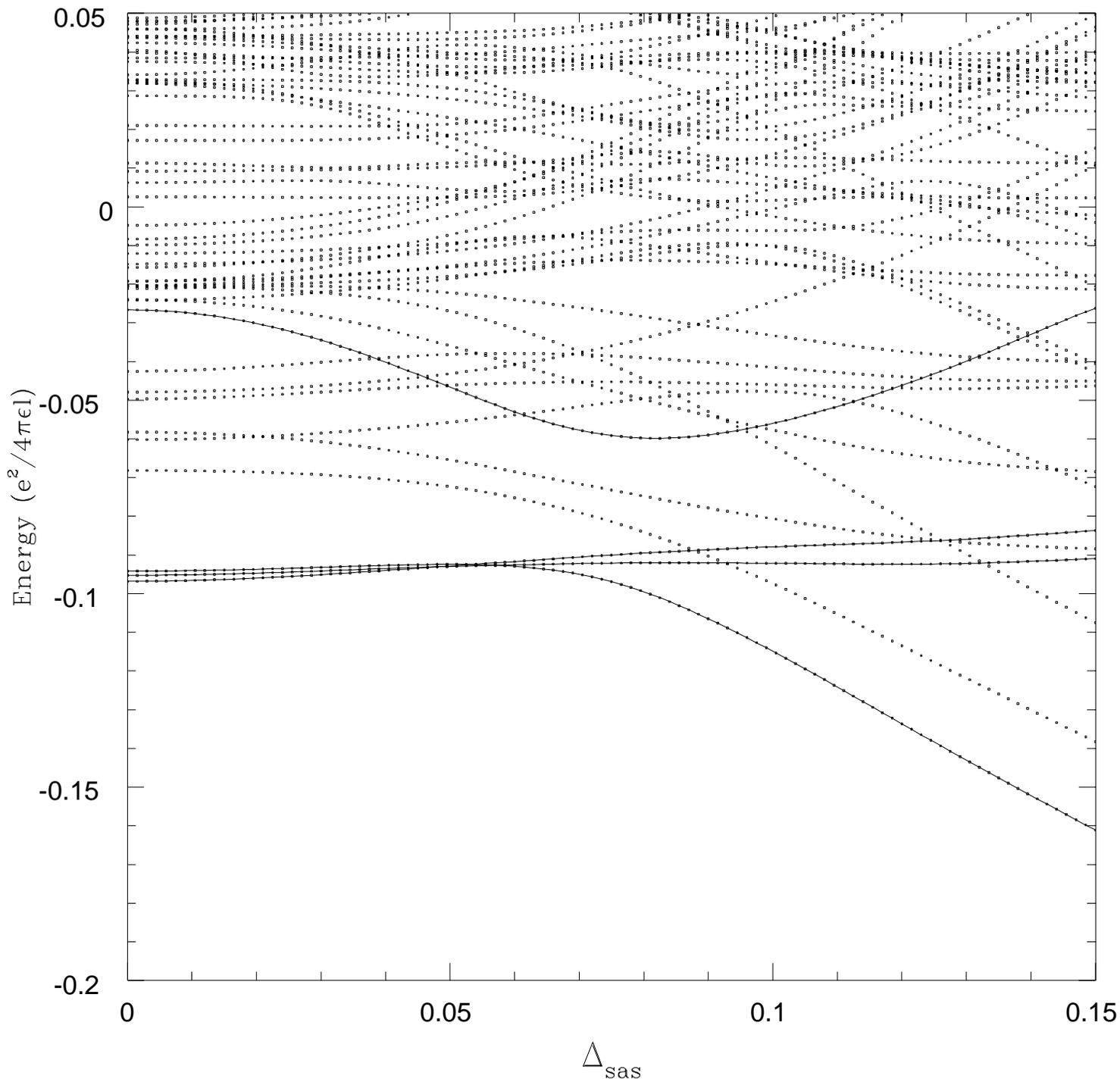
\end{document}